# Giant efficiency of long-range orbital torque in Co/Nb bilayers

Fufu Liu[1], Bokai Liang[1], Jie Xu[1], Chenglong Jia[1,2]†, Changjun Jiang[1,2]*

[1] Key Laboratory for Magnetism and Magnetic Materials, Ministry of Education, Lanzhou University, Lanzhou 730000, China

[2] Lanzhou Center for Theoretical Physics & Key Laboratory of Theoretical Physics of Gansu Province, Lanzhou University, Lanzhou 730000, China

Corresponding author. E-mail address: †cljia@lzu.edu.cn, *jiangchj@lzu.edu.cn

## ABSTRACT

We report unambiguously experimental evidence of a strong orbital current in Nb films with weak spin-orbit coupling via the spin-torque ferromagnetic resonance (ST-FMR) spectrum for Fe/Nb and Co/Nb bilayers. The sign change of the damping-like torque in Co/Nb demonstrates a large spin-orbit correlation and thus great efficiency of orbital torque in Co/Nb. By studying the efficiency as a function of the thickness of Nb sublayer, we reveal a long orbital diffusion length (~3.1 nm) of Nb. Further planar Hall resistance (PHE) measurements at positive and negative applying current confirm the nonlocal orbital transport in ferromagnetic-metal/Nb heterostructures.

Spin-related torques induced by spin currents have drawn considerable interest in spintronic over the decade [1-3]. Especially, spin Hall effect (SHE) prevails in heavy metal (HM) with strong spin-orbit coupling, where the resulting spin-orbit torque (SOT) has served as an efficient manipulation of magnetization in HM/ferromagnet (HM/FM) heterostructures [1, 4-7]. Recently, orbital current based on the flowing orbital angular momentum (OAM) [8] via orbital Hall effect (OHE) receives quite attention [9-12]. Like SHE, orbital current is generated along the perpendicular direction to the charge current. However, there is distinctive features of OHE compared to SHE. Theoretically, the OHE roots in orbital textures in momentum-space, which means that the OHE allows for the absence of spin-orbit coupling (SOC) due to the direct action of charge current on orbital degree of freedom [13], where charge current directly acts on spin degree of freedom through strong SOC that results in SHE. Moreover, orbital Hall conductivity (OHC) is much larger than spin Hall conductivity (SHC) in many transition metals, which could be estimated that the orbital torque (OT) efficiency can be comparable to or larger than spin torque efficiency [10]. Thus, the OT promotes the birth of orbitronics which holds great potential for future highly efficient magnetic devices and a complement to spintronics.

Many alternative materials can be regarded as the orbital source to induce OT, and theoretically found in multiorbital centrosymmetric systems such as transition metals [10], graphene [14], semiconductors [11] as well as two-dimensional transition metal dichalcogenides [15-17]. Experimentally, the OT was investigated in various systems, especially in thulium iron garnet TmIG/Pt/$CuO_x$ [18], FM/Cu/$Al_2O_3$ [19], Py/$CuO_x$ [20-21], etc. which all originates from the orbital Rashba effect (ORE) and modulates OT via layer design [22]. In these systems, Pt or oxidized Cu was utilized as an orbital source, it undoubtedly complicates the discussion of OT. Moreover, the intrinsic OHE was observed in Ta/Ni by evaluating the sign of damping-like torque similar to the sign of OHC in Ta [13]. However, the relatively larger SHC for Ta leads to slightly smaller OT efficiency in Ta/Ni system. Thus, materials selection without strong SOC for large OT efficiency is extremely necessary. The transition metal niobium (Nb) is characterized by weak SOC as well as large enough OHC ($\sigma_{OH}^{Nb}$) relative to SHC ($\sigma_{SH}^{Nb}$) and opposite sign of them [10,], which implies highly possible to observe obvious OHE via sign relation and generate expected OT acting on magnetization of FM [23]. Moreover, the large enough OHC is expected to be accompanied by a large OT efficiency.

Herein, we present experimental evidence confirming the existence of OHE in FM/Nb bilayers by the sign relation of damping-like torque according with OHC (or SHC) in Nb via spin-torque ferromagnetic resonance (ST-FMR). After fitting the FM dependence of ST-FMR spectra, the variety between the signs of damping-like torque for Fe/Nb and Co/Nb bilayers unambiguously proves the emergence of orbital torque. Furthermore, we support our conclusion that the sign of damping-like torque is examined by utilizing the method of planar Hall resistance (PHE) at positive and negative measuring current, which result coincides with ST-FMR and the theoretical calculations.

Generally, it is difficult to disentangle OT and ST due to identical properties. Thus, the total torque usually is ascribed to the combined effect of ST and OT, which includes following three cases: The first case corresponds to the same sign of $\sigma_{SH}$ and $\sigma_{OH}$ as shown in Figure 1(a). The total torque is enhanced by a synergistic effect of the OT and ST. In the second case, the sign of $\sigma_{SH}$ varies from $\sigma_{OH}$ as shown in the top of Figure 1(b), while the magnitude of spin Hall contribution ($A_{\mathrm{SH}}$) is larger than orbital Hall contribution ($A_{\mathrm{OH}}$), the total torque is dominated by ST. In the final case, the sign of $\sigma_{SH}$ varies from $\sigma_{OH}$ in the bottom of Figure 1(b), and the $A_{\mathrm{SH}}$ is smaller than $A_{\mathrm{OH}}$ that provides a great chance to disentangle the OT from the ST. Given all this, the 4d transition metal Nb is an excellent candidate to observe OT where $\sigma_{OH}$ is opposite sign and much larger than $\sigma_{SH}$ [10]. Specifically speaking, as shown in Figure 1(c), the orbital current and spin current are generated by OHE and SHE and injected into FM, respectively. Next the injected orbital current is converted to spin via SOC of the FM. A crucial conversion ratio $\eta_{FM}$ (spin-orbit correlation) [24] is used to describe how much spin in FM is induced by the orbital current injected from the Nb. For most FMs, the sign of $\eta_{FM}$ is positive, such as Fe, Co, Ni, etc. However, the small $\eta_{FM}$ tends to induce ST, while the large $\eta_{FM}$ leads to OT, as demonstrated in Figure 1(c). Based on the previous theoretical[13], we summary the total torque of FM/Nb systems in Figure 1(d). Obviously, the Co/Nb bilayer is a good candidate to clarify the OT, given that the sign of the effective Hall conductivity ($\sigma_{SH}^{Nb} + \eta_{FM}\sigma_{OH}^{Nb}$) obtained from torque coincides with the sign of $\sigma_{OH}^{Nb}$. The FM/Nb thin films were deposited on MgO substrate by magnetron sputtering at room temperature. In order to clarify the physical mechanism of OT in experiments, four different types of bilayers were prepared: Fe($t_{Fe}$)/Nb(6 nm), Co($t_{Co}$)/Nb(6 nm). Fe(8 nm)/Nb($t_{Nb}$) and Co(8 nm)/Nb($t_{Nb}$). For comparison, the samples Pt(6 nm)/Co($t_{Co}$) and Pt(6 nm)/Fe($t_{Fe}$) were grown on MgO substrate as well. Moreover, Fe(2 nm)/Nb(8 nm) and Co(2 nm)/Nb(8 nm) are prepared to conduct the PHE measurements.

To measure the OT efficiency, we exploit the spin-torque ferromagnetic resonance (ST-FMR) measurement as shown in Figure 2(a), the ST-FMR technique is employed because it is a well-established method to determine the charge-spin conversion efficiency as well as the type, number and efficiency of torques [1, 25-28]. In this measurement, a radio frequency (RF) microwave current is applied along the x-direction, this RF current will induce the resulting torque and cause the FM magnetization process, which contributes to the ST-FMR resonance signal. The typical ST-FMR signal for Fe(8 nm)/Nb(6 nm) is shown in Figure 2(b), which could be fitted by symmetric ($V_s$) and antisymmetric component ($V_a$). Generally, the $V_s$ and $V_a$ corresponds to different torques, where the $V_s$ arises solely from the damping-like torque ($\boldsymbol{\tau}_{\mathbf{DL}} \propto \boldsymbol{m} \times (\boldsymbol{y} \times \boldsymbol{m})$) and $V_a$ originates jointly from the field-like torque ($\boldsymbol{\tau}_{\mathbf{FL}} \propto \boldsymbol{m} \times \boldsymbol{y}$) and current-induced Oersted field, respectively [13]. Furthermore, the quantitative conversion efficiency ξ can be determined by the ratio of $V_s$ and $V_a$ based on the following equation [1, 13, 29-30]:

$$\xi = \frac{V_S}{V_a}\frac{e\mu_0 M_S t_{FM} d_{Nb}}{\hbar}[1 + (M_S/H_r)]^{1/2}, \quad (1)$$

where $t_{FM}$ and $d_{Nb}$ are the thickness of FM and Nb sublayers, respectively. $H_r$ is the resonance field. $M_s$ is the effective saturation magnetization that can be obtained by Kittel equation: $(2\pi f)/\gamma = \sqrt{[H_r(H_r + M_S)]}$ with $\gamma$ being the gyromagnetic ratio.

Analogous to torque components of SOT, the $\boldsymbol{\tau}_{\mathbf{DL}}$ and $\boldsymbol{\tau}_{\mathbf{FL}}$ components of OT can be generated in adjacent magnetic layer and separately determined via ST-FMR. In the present study, we are more interested in the sign of $\boldsymbol{\tau}_{\mathbf{DL}}$, which could reflect the sign of SHC and/or OHC. As shown in Figure 2(b), a negative $V_s$ is observed for Fe(8 nm)/Nb(6 nm), which means the existence of $\boldsymbol{\tau}_{\mathbf{DL}}$. Obviously, the sign of $V_s$ is consistent with the expected sign of $\sigma_{SH}^{Nb}$ and opposite to the $\sigma_{OH}^{Nb}$. The sign of $\boldsymbol{\tau}_{\mathbf{DL}}$ for Fe/Nb could be further determined by the FM layer thickness dependence of the ST-FMR resonance signal $V_{dc}$ [24] as shown in Figure 2(c). The experimentally measured conversion efficiency ξ as a function of $t_{FM}$ provides a route to separately evaluate the damping-like ($\xi_{DL}$) and field-like ($\xi_{FL}$) torque efficiencies as [13, 20]

$$\frac{1}{\xi} = \frac{1}{\xi_{DL}}\left(1 + \frac{\hbar}{e}\frac{\xi_{FL}}{\mu_0 M_s t_{FM} d_{Nb}}\right) \quad (2)$$

where the intercept implies the sign of $\xi_{DL}$. As shown in Figure 2(c), the sign of the intercept is in accord with the symmetric component $V_s$ in Figure 2(b). Hence, it is apparent that the sign of $\xi_{DL}$ of Fe/Nb bilayer is consistent with the sign of $\sigma_{SH}^{Nb}$. However, we have a sign reversal of the symmetric component $V_s$ in ST-FMR signal for Co(8 nm)/Nb(6 nm), i.e., a positive $V_s$ as shown in Figure 2(d). Consequently, the FM layer thickness dependence of efficiency for Co/Nb bilayers [24] gives a positive $\xi_{DL}$ as plotted in Figure 2(e), which coincides with the expected sign of $\sigma_{OH}^{Nb}$ but opposite to the sign of $\sigma_{SH}^{Nb}$. On the other hand, for the Pt/Fe and Pt/Co control sample as shown in Figures 3, the heavy metal Pt, is acknowledged as a strong spin source material [1]. From the experimental result, the positive sign of $V_s$ is consistent with the sign of $\sigma_{SH}^{Pt}$, and further accords with the sign of $\xi_{DL}$. According to schematic illustration for the mechanism of the orbital torque in Figure 1(c) and theoretical calculations [24], the total torque in Co/Nb includes conventional SOT arising from spin Hall contribution ($\sigma_{SH}^{Nb}$) and OT attributing to orbital Hall contribution ($\sigma_{OH}^{Nb}$). When the spin Hall contribution is much less than orbital Hall contribution, the sign of the total torque is same as the sign of $\sigma_{OH}^{Nb}$. In this case, as for Co/Nb bilayer, the sign of OT depends on the $\sigma_{OH}^{Nb}$ and spin-orbit correlation $\langle \mathbf{L}\cdot\mathbf{S}\rangle^{Co}$ [13]. The fact from theoretical calculation is that the orbital Hall effect contribution is larger than the spin Hall one in magnitude of Co/Nb, the sign of total torque is in accord with the expected sign of $\sigma_{OH}^{Nb}$, i.e., the orbital torque.

Furthermore, to verify the source of OT, we perform the ST-FMR measurement for Co(8 nm)/Cu(t)/Nb(6 nm), Figure S4 shows the typical ST-FMR spectra. The positive $V_s$ means the existence of OT even if inserting the Cu layer. Cu layer thickness dependence of efficiency ξ is plotted in Figure 4(a). Enough efficiency excludes the interface effect at Co/Nb interface and bulk OHE is responsible for the OT. Moreover, to obtain the orbital diffusion length of Nb, a series of ST-FMR samples with different Nb thicknesses of Co(and Fe, t = 8 nm)/Nb($d$ = 6-15 nm) was fabricated, and the conversion efficiency ξ was characterized by Eq. (1) and summarized in Figure 3(a), the Nb thickness dependence of ξ match with $\xi = \xi_{\infty}[1 - \text{sech}\,(d/\lambda_s)]$ [31], where ξ is the measured conversion efficiency at different Nb thickness and $\xi_{\infty}$ is the conversion efficiency at infinite Nb thickness. From the fitting, $\lambda_s$ of Nb is quantitatively determined to be 3.1 nm. This point sets it apart from conventional heavy metal, such as Pt, where spin Hall contribution dominates and shows a slightly small spin diffusion length (~ 1.5 nm) [31].

Furthermore, to verify the sign relation of damping-like torque in FM/Nb bilayers, we perform the method of PHE at positive and negative measuring current, which has been regarded as a reliable and effective method to measure charge-to-spin conversion [32-34]. In this measurement, the damping-like torque induces the out-of-plane effective field $H_p$, while the field-like torque induces the in-plane effective field $H_T$ as shown in Figure 4(a). Here we are just concerned with the sign of $H_p$. Figure 4(b) and (c) exhibits the typical magnetic angle dependence of planar Hall resistance $R_H$ as well as the corresponding $R_{DH}$ curve at $I = \pm$ 0.9 mA for Fe(2 nm)/Nb (8 nm), and $I = \pm$ 1.0 mA for Co (2 nm)/Nb(8 nm), respectively (see Supplemental Material for other currents [24]). Note that the effective fields $H_p$ and $H_T$ can be extracted from the resistance difference $R_{DH}$ and charactered by following equation [32-34]:

$$R_{DH}(I, \varphi) = 2R_H \frac{(H_T + H_{Oe})}{H_{ext}} (cos\varphi + cos3\varphi) + 2\frac{dR_{AHE}}{dH_{perp}} H_P cos\varphi + C \qquad (3)$$

where C is the resistance offset, $H_{Oe}$ is the Oersted effective field, $\varphi$ is defined in Figure 4(a), $H_{perp}$ is the applied out-of-plane magnetic field, $\frac{dR_{AHE}}{dH_{perp}}$ is the slope of $R_{AHE}$ vs $H_{perp}$ [24]. After determining relevant parameters in Eq. (3), the out-of-plane effective field $H_p$ is obtained by fitting the $\varphi$ dependence of $R_{DH}$. Figure 4(d) shows the linear fitting $H_p$ relative to the current for Fe/Nb and Co/Nb, respectively. The sign of effective conversion efficiency can be quantitatively determined by [32]:$\xi \sim \frac{H_p}{J_{NM}}$, where $J_{NM}$ is the density of current. Obviously, the sign of $\xi$ depends on the slope of $H_p$ versus current $I$. Any difference in sign of slope for $H_p$ vs $I$ between Fe/Nb and Co/Nb would suggest an existence of orbital torque, which is consistent with ST-FMR result and the theoretical calculations.

In conclusion, the generation of orbital current and orbital torque is experimentally confirmed based on orbital Hall effect present in FM/Nb systems measured by spin-torque ferromagnetic resonance. After fitting the FM thicknesses dependence of the ST-FMR resonance signal $V_{dc}$, we characterize the sign of $\xi_{DL}$ for Fe/Nb and Co/Nb is positive, which is consistent with the spin Hall conductivity and orbital Hall conductivity of Nb from theoretical calculations, respectively. This result can be further verified by the method of planar Hall resistance. This strategy of generating orbital current and orbital torque by experimental method provide a route of enabling applications in spintronic and orbitronic devices by orbital engineering.

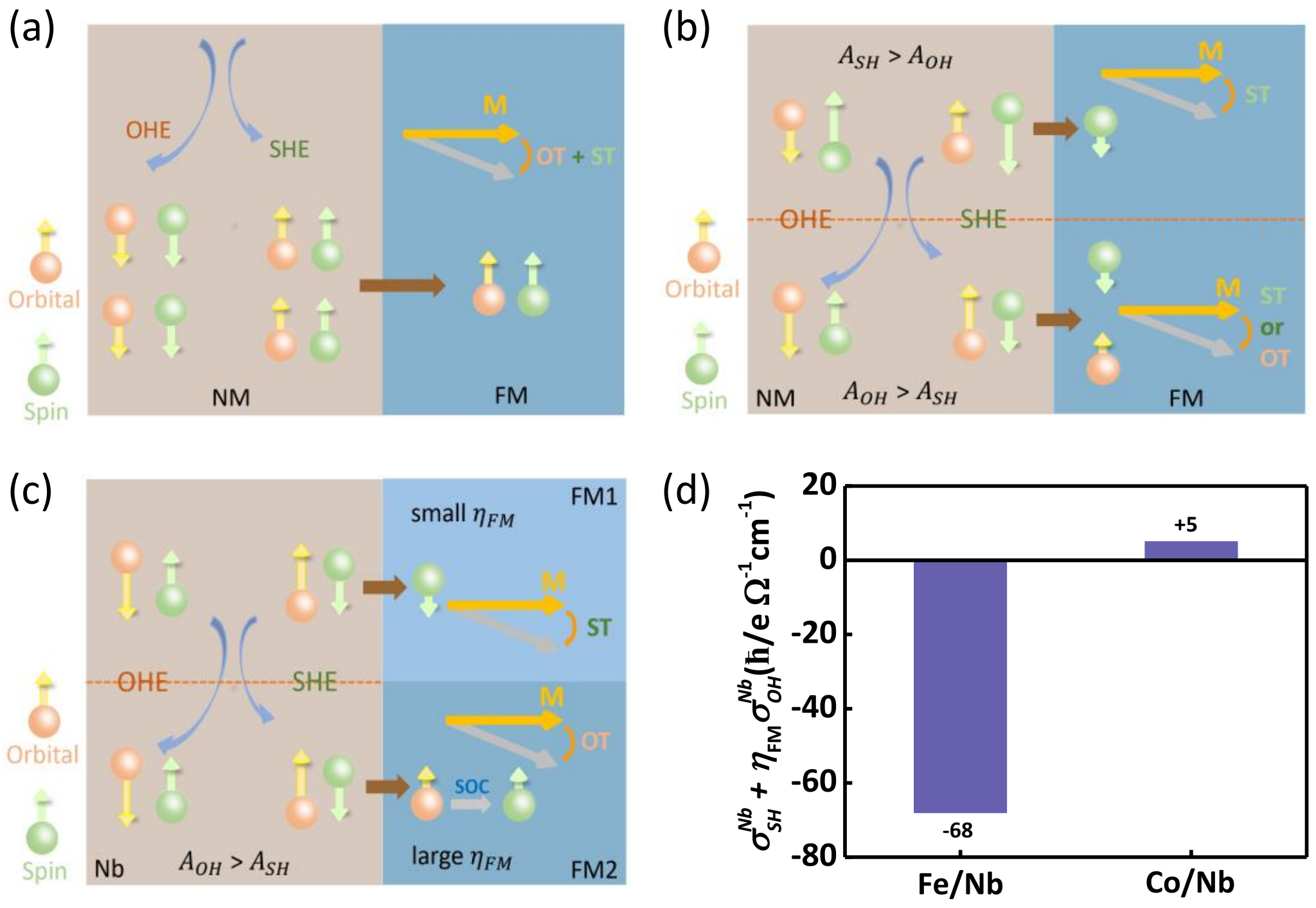

**Figure 1**. (a) The same sign for $\sigma_{SH}$ and $\sigma_{OH}$, where the total torque includes the sum of OT and ST. (b) The opposite sign of $\sigma_{SH}$ and $\sigma_{OH}$. The total torque could be either ST or OT. (c) Schematic illustration for the mechanism of the orbital torque in FM/Nb. The orbital current generated by OHE in the Nb is injected into FM. Next the injected orbital current is converted to spin through SOC of the FM. The resulting torque exerted by this spin exerts and acts on the magnetization of FM, which is referred to as orbital torque (OT). (d) Theoretical calculation for $\sigma_{SH}^{Nb} + \eta_{FM}\sigma_{OH}^{Nb}$ of FM/Nb bilayers (FM = Fe and Co). Where $\sigma_{SH}^{Nb}$, $\sigma_{OH}^{Nb}$ and $\eta_{FM}$ are spin Hall conductivity, orbital Hall conductivity of Nb and orbital-to-spin conversion efficiency of FM, respectively.

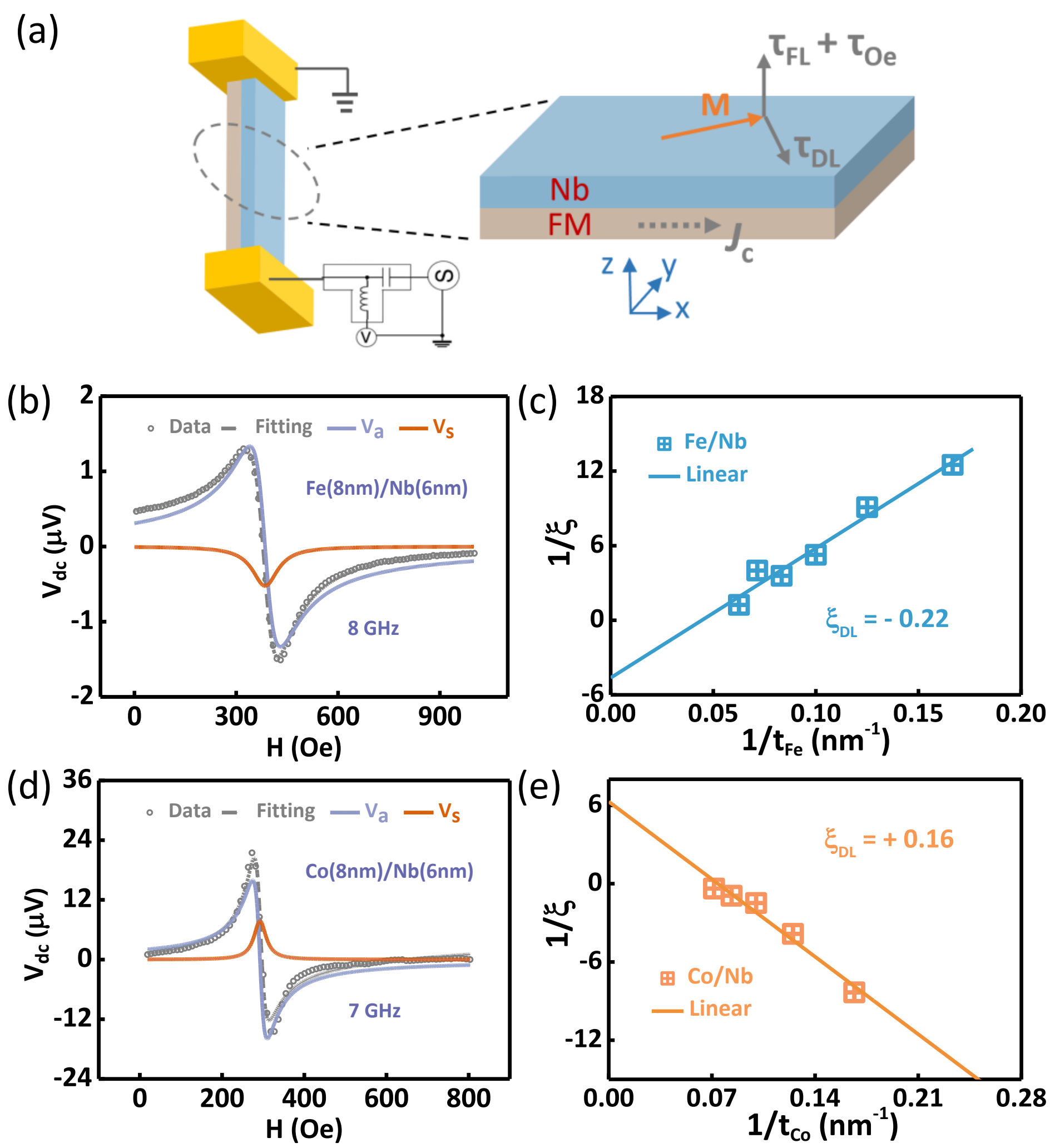

**Figure 2**. (a) Schematic of the circuit for ST-FMR measurements. ST-FMR voltage signal $V_{dc}$ for (b) MgO/Fe(8 nm)/Nb(6 nm) at 8 GHz, (d) MgO/Co(8 nm)/Nb(6 nm) at 7 GHz. The inverse of the conversion efficiency $1/\xi$ as a function of $1/t_{FM}$ for (c) Fe/Nb bilayer and (e) Co/Nb bilayer. The solid lines are the linear fit to the experimental data (grids).

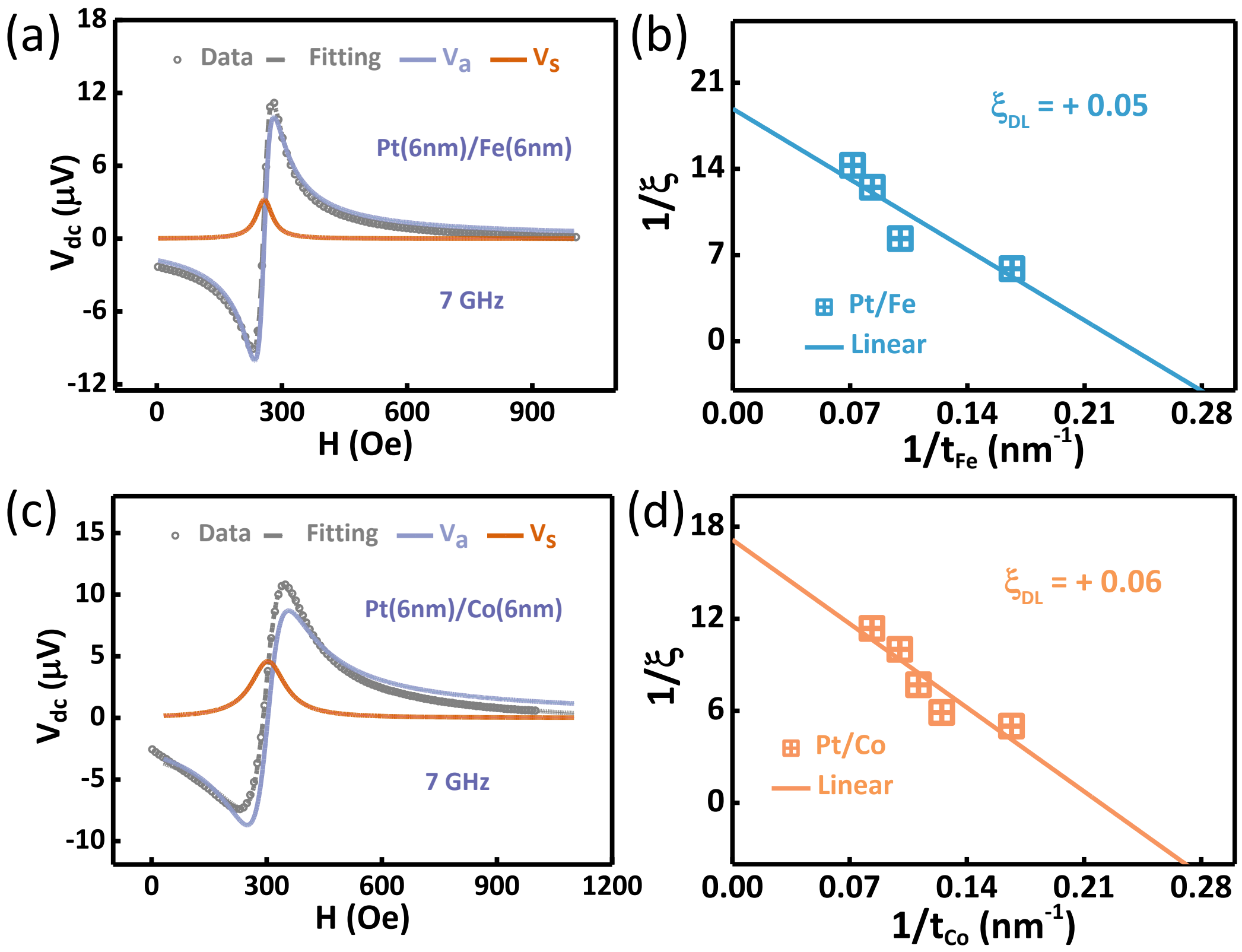

**Figure 3**. ST-FMR voltage signal $V_{dc}$ for (a) MgO/Pt(6 nm)/Fe(8 nm) at 7 GHz, (c) MgO/Pt(6 nm)/Co(8 nm) at 7 GHz. The inverse of the conversion efficiency $1/\xi$ as a function of $1/t_{FM}$ for (b) Pt/Fe bilayer and (d) Pt/Co bilayer. The girds are the experimental data, and the solid line is the linear fit to the data.

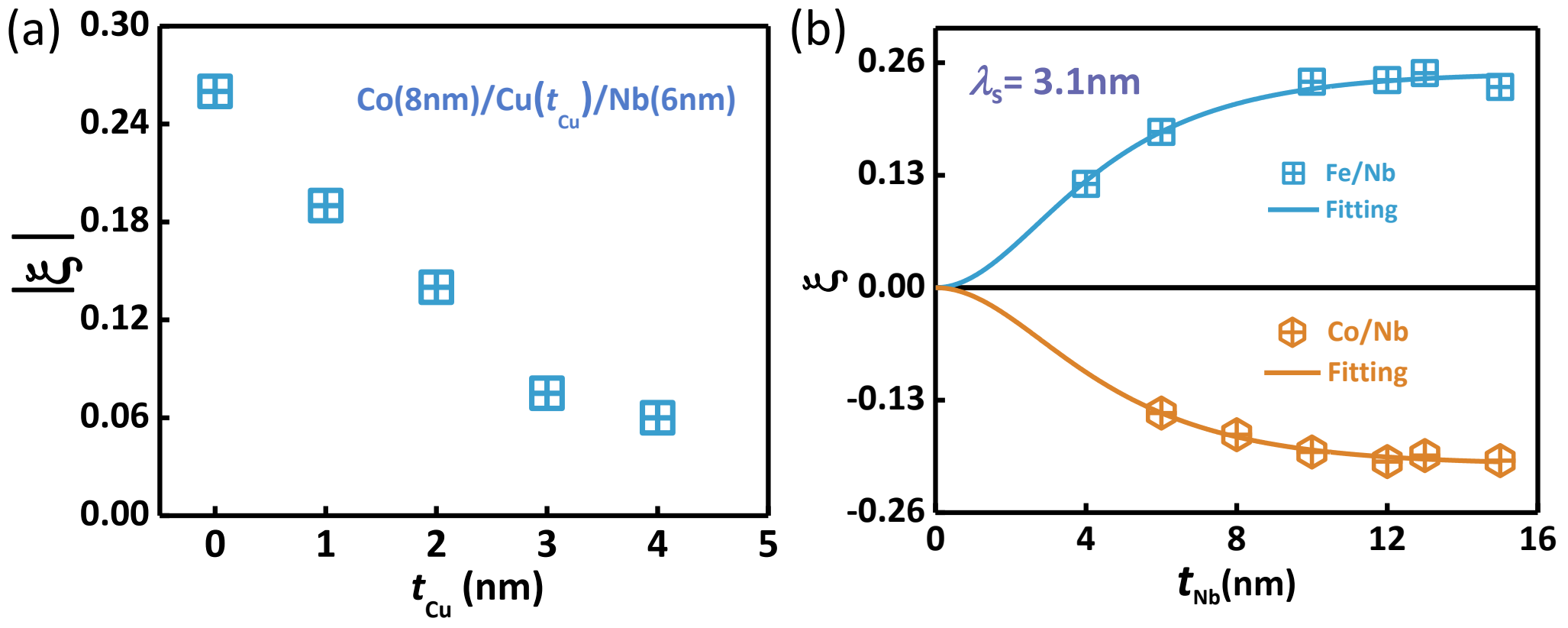

**Figure 4**. (a) Cu-layer thickness $t_{Cu}$ dependence of efficiency for Co(8 nm)/Cu($t_{Cu}$)/Nb(6 nm). (b) The conversion efficiency ξ (orange squares) as a function of Nb thickness for Co (8 nm)/Nb(t) as well as Fe (8 nm)/Nb(t) systems and a fit (solid orange line).

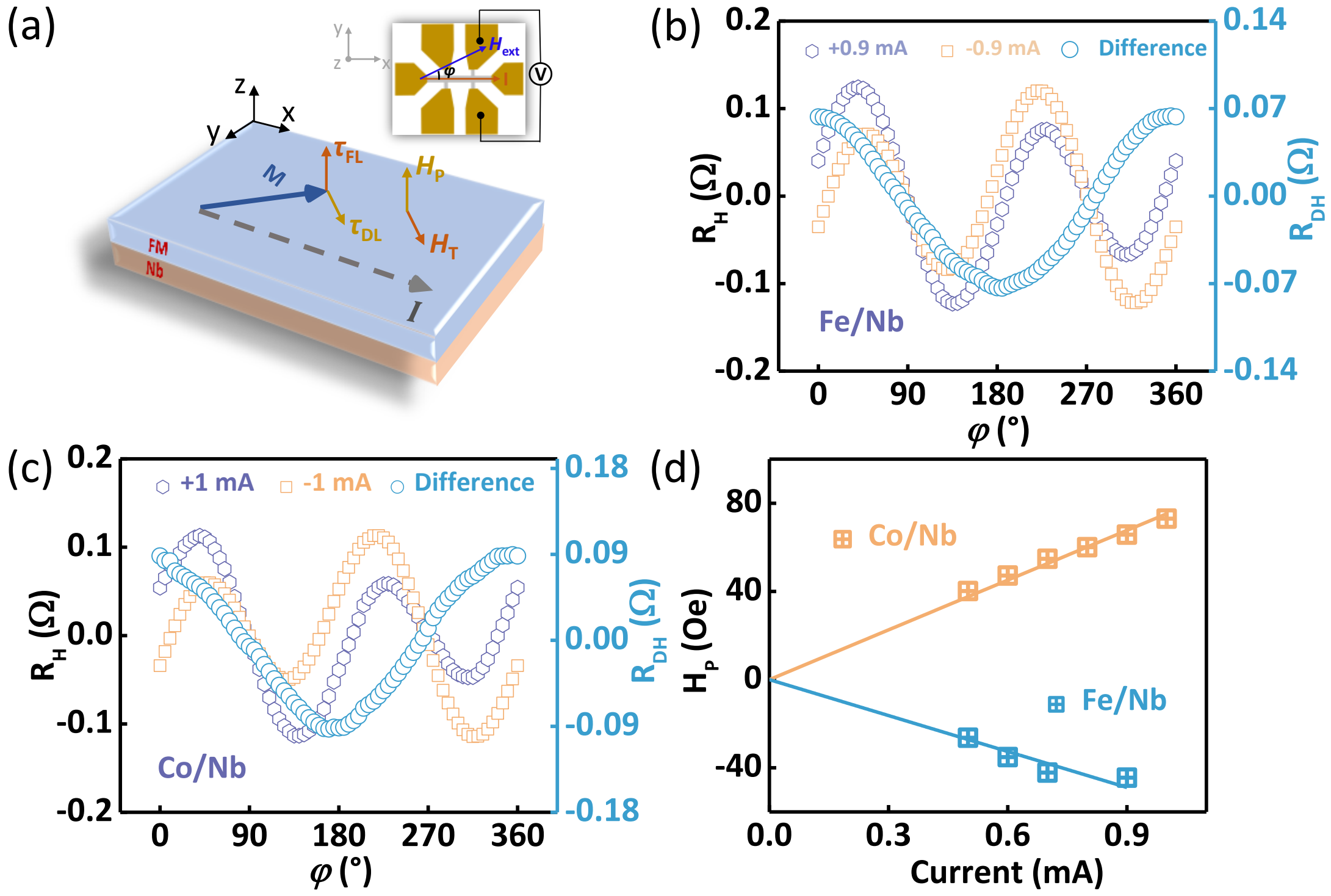

**Figure 5**. (a) Current-induced effective fields and schematic illustration of the planar Hall resistance measurement srtup. The external magnetic field angle $\varphi$ dependence of planar Hall resistance $R_H$ of (b) Fe(2 nm)/Nb(8 nm) and (c) Co(2 nm)/Nb(8 nm). The right axis of (b) and (c) represents the difference of the Hall resistances $R_{DH}$ at positive and negative currents. (d) The variety of $H_p$ with the applied current for Fe(2 nm)/Nb(8 nm) and Co(2 nm)/Nb(8 nm).